\documentclass{article}

\usepackage{PRIMEarxiv}

\usepackage[utf8]{inputenc} % allow utf-8 input
\usepackage[T1]{fontenc}    % use 8-bit T1 fonts
\usepackage{hyperref}       % hyperlinks
\usepackage{url}            % simple URL typesetting
\usepackage{booktabs}       % professional-quality tables
\usepackage{amsfonts}       % blackboard math symbols
\usepackage{nicefrac}       % compact symbols for 1/2, etc.
\usepackage{microtype}      % microtypography
\usepackage{lipsum}
\usepackage{fancyhdr}       % header
\usepackage{graphicx}       % graphics
\usepackage{multirow}
\graphicspath{{media/}}     % organize your images and other figures under media/ folder

\title{Are AI Agents interacting with Online Ads?
}

\author{
  Andreas Stöckl \\
  Digital Media Lab \\
  University of Applied Sciences Upper Austria \\
  \texttt{andreas.stoeckl@fh-hagenberg.at} \\
   \And
  Joel Nitu \\
  Digital Media Lab \\
  University of Applied Sciences Upper Austria \\
  \texttt{joel.nitu@fh-hagenberg.at}  \\
}

\begin{document}
\maketitle

\begin{abstract}
As AI-driven agents become increasingly integrated into the digital ecosystem, they reshape how online advertising is perceived and processed. Particularly in the travel and hotel booking sector, these autonomous systems influence the effectiveness of traditional advertising formats. While visual cues and emotional appeals sway human users, AI agents prioritize structured data such as price, availability, and specifications. This study examines how different AI agents interact with online advertising, whether they incorporate ads into their decision-making processes, and which ad formats prove most effective. We analyze interaction patterns, click behavior, and decision-making strategies through experiments with multimodal language models such as OpenAI GPT-4o, Anthropic Claude 3.7 Sonnet, and Google Gemini 2.0 Flash. Our findings reveal that AI agents neither ignore nor systematically avoid advertisements but instead favor certain features—particularly keywords and structured data. These insights have significant implications for the future design of advertising strategies in AI-dominated digital environments.
\end{abstract}

% keywords can be removed
\keywords{Autonomous agents \and Intelligent agents \and Multimodal language models \and Online advertising \and User interaction}

\section{Introduction}
As artificial intelligence (AI) agents become increasingly sophisticated, they are poised to reshape the digital ecosystem significantly. These autonomous agents—capable of navigating websites, interpreting content, and making decisions for human users—are emerging as novel intermediaries in online searches and e-commerce transactions. In the travel sector, for instance, an AI-powered booking assistant could sift through vast numbers of hotel deals and flight options far more comprehensively than a human user, thereby potentially altering the effectiveness and reach of traditional online advertising. Industry projections suggest that by 2026, up to 25\% fewer searches will be performed on conventional search engines in favor of AI-driven assistants \cite{browser_use2024}. Consequently, a substantial reduction in traditional exposure of advertisements is anticipated, prompting urgent questions about how best to design and deliver promotional content in an AI-mediated environment.

Notably, AI agents differ from human users' interaction with online ads. Unlike human consumers, who might be swayed by visual cues, emotional appeals, or brand messaging, AI agents prioritize structured, factual data such as price, specifications, or availability. This selective attention can render conventional advertising formats—banner ads, pop-ups, and branded content—less persuasive or even irrelevant when the ultimate decision-maker is a machine \cite{nakano2021webgpt, wang2024gui}. On the one hand, these agents might be more rational and objective than humans. On the other, they can also be susceptible to highly technical manipulations, including adversarial pop-ups designed to exploit vulnerabilities in their vision-language models \cite{zhang2024large}. At the same time, fraud analyses indicate that a significant proportion of existing online ad clicks already originate from non-human traffic, often in the form of malicious bots \cite{hoscilowicz2024clickagent}. Thus, the rise of benevolent AI agents raises parallel concerns about how to measure and validate advertising effectiveness in an environment heavily populated by automated interactions.

This paper investigates how AI agents respond to—and are influenced by—online advertising, focusing on the domain of hotel and travel booking platforms. We examine whether these agents incorporate ads as meaningful information sources, how different ad formats (e.g., banners, native advertising) affect agent decision-making, and how industry stakeholders might adapt advertising strategies to remain effective.

\section{Related Work}
\label{sec:relatedwork}

Recent advances in large language models (LLMs) have enabled a new class of agents that can control web browsers and graphical user interfaces (GUIs) through natural language \cite{zhang2024large}. Early efforts, such as WebGPT, fine-tuned GPT-3 to navigate a text-based browser for question answering \cite{nakano2021webgpt}. These demonstrated the potential of LLMs to interpret instructions and retrieve information via web actions. Today’s systems go further—using multimodal LLMs to not only read web content but also click buttons, fill forms, and execute multistep tasks in a browser, just as a human would \cite{zhang2024large, wang2024gui}. We initially compare three prominent approaches in this domain: OpenAI’s Operator, Anthropic’s Claude Computer Use, and an open-source Browser Use agent \cite{browser_use2024}. Each represents a distinct design in how LLMs are leveraged for browser automation.

\subsection{OpenAI Operator – LLM-Based Browser Agent}

Operator is OpenAI’s first ``agent'' that autonomously operates a web browser via a chat interface. It is powered by a new Computer-Using Agent (CUA) model derived from GPT-4 \cite{achiam2023gpt}. This model combines GPT-4o’s vision capabilities with advanced reasoning techniques (enhanced through reinforcement learning) to understand web page screenshots and interact with on-screen elements. Operator runs in a cloud-hosted Chrome environment, iterating through cycles of perception (reading the page), planning, and action execution until the user’s task is completed\footnote{\url{https://www.infoq.com/news/2025/02/openai-operator-release/}}. The agent perceives the webpage primarily through vision—interpreting rendered content pixel‑by‑pixel—and can simulate human-like actions (clicking links, typing text, scrolling) guided by the LLM’s decisions. 

The core GPT-4o model in Operator serves as the vision interpreter and the task planner. By feeding it browser screenshots, the agent lets the LLM ``see'' the state of the page and respond with the following action. OpenAI reports that this vision-and-action loop, combined with instruction-following fine-tuning and reinforcement learning, enables a robust understanding of GUI components like buttons, menus, and form fields. The LLM effectively outputs a sequence of commands or high-level actions executed in the browser, achieving fully automated web-based task completion.

\subsection{Anthropic Claude – Computer Use Feature}

Anthropic’s Claude 3.5 introduced a ``computer use'' mode that enables the LLM to control a desktop interface in a human-like manner. This approach treats the screen as the primary observation: Claude receives periodic screenshots of the user’s computer display and issues mouse and keyboard actions. Under the hood, Anthropic trained a version of Claude (dubbed Sonnet) with a focus on GUI manipulation skills\footnote{\url{https://www.anthropic.com/news/developing-computer-use}}. The model learned to interpret GUI imagery and precisely move a cursor by ``counting pixels'' to target interface elements. Notably, only a small set of simple software (e.g., a calculator app, text editor) was used in supervised training without internet access for safety. Despite this limited training, the enhanced Claude can generalize surprisingly well–breaking down novel tasks into action sequences and even self-correcting if an attempt fails.

Claude’s computer-use capability is built into the Claude LLM itself rather than relying on an external tool API. The model accepts an image of the current screen (or some encoded form of it) and the user’s last instruction, then directly outputs a low-level action (e.g., ``move mouse and click'') in a single step. Because reasoning and perception are unified in a single model, Claude takes a natural-language instruction and internally devises the exact sequence of GUI actions needed to carry it out. Anthropic's research emphasizes the importance of spatial reasoning in this setup, which is why the focus is on pixel-precise cursor movements learned during fine-tuning. The result is an end-to-end LLM-based agent that can see the state of the UI and act accordingly without requiring intermediate symbolic representations like HTML trees.

\subsection{Browser Use: Open-Source DOM-Focused Agent}

An alternative approach to LLM-based browsing is exemplified by Browser Use, an open-source project that acts as an ``Operator'' analog for any LLM. Instead of training a custom multimodal model, this system connects a standard web browser to an LLM via a programmatic interface. The agent obtains a structured representation of the web page (the entire DOM tree, including text and element attributes) and feeds this as context to the LLM, then outputs an action. The action is executed in the browser through automation (e.g., using Playwright\footnote{\url{https://playwright.dev/}} or similar), and the cycle repeats. Browser Use leverages the browser’s Document Object Model as a parser rather than requiring the LLM to interpret pixels visually. This design grants the agent access to all page content, even if hidden or off-screen, by directly reading the HTML/XML elements \cite{hoscilowicz2024clickagent}. Prior work has shown that providing structured hints (like DOM IDs, text labels, or OCR-extracted text) can aid LLMs in identifying targets on the interface.

LLM Agnosticism: A key feature of the Browser Use approach is model flexibility. It is designed to work with any large language model through prompting rather than a single fine-tuned brain. In practice, developers can plug in an API for GPT-4 \cite{achiam2023gpt}, Mistral \cite{jiang2023mistral}, Claude \cite{anthropic2024claude}, Google’s Gemini \cite{team2023gemini}, an open-source Llama3 \cite{grattafiori2024llama}, etc., and the system will prompt that model with the page’s DOM description and the user’s instruction. This aligns with Microsoft’s vision of ``turning any LLM into a computer-use agent,'' as seen with OmniParser \cite{lu2024omniparser} and OmniTool research. Indeed, tools like OmniParser can be seen as complementary: OmniParser converts raw screenshots into a structured list of UI elements, which could then be fed to an LLM for action planning. The Browser Use agent effectively skips the vision step on web pages by using the DOM as a readily available structured representation. This yields a highly extensible framework—as new, more capable LLMs emerge, they can be swapped in to enhance the agent’s reasoning or understanding abilities immediately.

\subsection{AI Agents and Online Advertising}

Research on AI agents’ interaction with online advertising spans multiple perspectives, ranging from empirical investigations of agent vulnerabilities to market analyses that predict broader shifts in digital advertising models:

\begin{itemize}
\item Pop-up Vulnerabilities. 

Zhang et al. \cite{zhang2024attacking}  explored how vision-language AI agents can be manipulated via pop-up advertisements in web-browsing tasks. Their study found that agents lacking robust ad-detection heuristics exhibited an 86\% click-through rate on deceptive pop-ups, severely undermining their ability to accomplish primary tasks.

\item Advertising Model Disruption. 

Curley\footnote{\url{https://www.carbon6.io/blog/ai-agents-amazon-advertising-ppc}} provided an industry-focused analysis of how AI agents reshape pay-per-click (PPC) advertising, concluding that agents often bypass both sponsored listings and banner ads. The report forecasts a significant disruption to traditional PPC models as merchants and advertisers struggle to position themselves within the agents’ streamlined decision processes.

\item Machine-Readable Marketing. 

Ketchell\footnote{\url{https://www.leaddigital.com/blog/marketing-to-machines/}} argued that marketing must evolve toward ``machine-to-machine'' interactions. In contrast to humans, AI agents place little value on emotional or visual appeal; instead, they rely on structured data feeds and APIs. This approach allows for ``API-driven marketing,'' wherein advertisers pay for preferential treatment within an agent’s ranking algorithms rather than via conventional display ads.

\item Market Projections. 

Gartner’s market study\footnote{\url{https://searchengineland.com/search-engine-traffic-2026-prediction-437650}} underscored the broader shift away from standard search engines to AI-driven chatbots and assistants. The study projects a notable decline in traffic to traditional search result pages, which could, in turn, reduce the visibility of existing ad formats. This trend indicates a need for advertisers to tailor their strategies specifically for AI-mediated user journeys.

\item Travel Industry Implications. 

Watts\footnote{\url{https://www.hotelnewsresource.com/article134429.html}} explicitly focused on how AI agents may redefine competitive dynamics in the travel and hospitality sectors. By evaluating countless hotel and flight listings, AI travel assistants tend to ignore brand-driven advertisements, prompting travel companies and online travel agencies (OTAs) to rethink their reliance on banner ads and sponsored placements.

\item Fraudulent Traffic Insights. 

Juniper Research\footnote{\url{https://trustedclicks.ai/top-10-types-of-click-fraud/}} examined the prevalence of fraudulent clicks in online advertising. While this report primarily addresses malicious or fraudulent bots, it highlights how non-human traffic already inflates ad engagement metrics. By extension, it raises crucial questions about measuring advertising impact in a landscape where an increasing share of ``users'' could be automated agents.
\end{itemize}

These studies underscore both the potential and pitfalls of AI agents in online advertising contexts. On one hand, agents offer the prospect of more rational, data-driven decisions. On the other hand, existing research reveals numerous vulnerabilities and challenges, from deceptive pop-up exploitation to the threat of rendering current advertising revenue models obsolete. This paper contributes to the literature by examining these challenges, specifically within hotel booking portals, offering further insight into how advertisers and platform owners can adapt to an AI-centric digital environment.

\section{Experimental Setup}

\begin{figure*}[htbp]
\centerline{\includegraphics[width=0.8\textwidth]{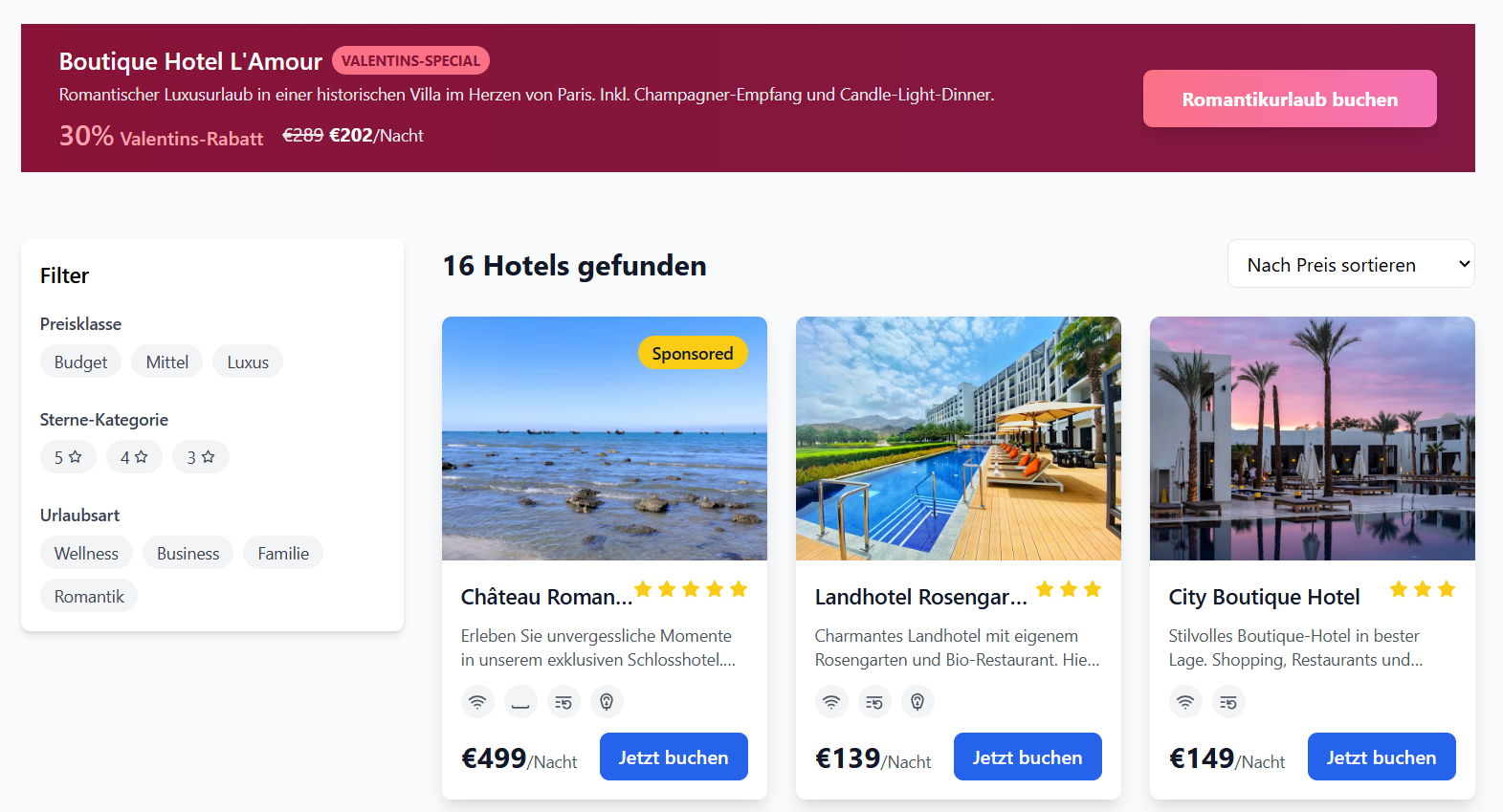}}
\caption{The experimental environment utilized for testing.}
\label{fig:fig1}
\end{figure*}

Figure~\ref{fig:fig1} illustrates our experimental environment, which consisted of a custom-developed hotel booking platform. This platform was designed to replicate real-world travel websites, incorporating elements such as a banner area, filtering tools, and a grid layout showcasing various hotel listings along with their respective booking detail pages.

The platform featured dynamic hotel listings, price comparisons, availability filters, and embedded online advertisements in various formats, including:
\begin{itemize}
\item Banner ads
\item Native advertisements designed to blend with organic content.
\end{itemize}

The prototype of the hotel booking portal was developed using the React framework, a widely adopted JavaScript library for building interactive user interfaces \cite{fedosejev2015react}. React was chosen due to its component-based architecture, which enhances modularity and reusability, facilitating rapid prototyping and iterative design processes \cite{lee2004component}. The application features a dynamic front-end, where users can search for hotels, filter results based on preferences, and complete bookings through an integrated form.

The contents are artificially generated hotel descriptions and attributes in German using free images from Unsplash\footnote{\url{https://unsplash.com/}}.

To investigate the interaction between AI agents and online advertising in the context of hotel booking, we conducted a series of experiments using the systems Browser Use\footnote{\url{https://github.com/browser-use/browser-use}} \cite{browser_use2024} and OpenAI Operator\footnote{\url{https://openai.com/index/computer-using-agent/}}, state-of-the-art autonomous agents designed to perform web-based tasks. Unlike traditional chatbot-based AI models, they actively engage with websites by navigating interfaces, filling out forms, and completing transactions. This capability enables us to examine how AI-driven agents perceive and interact with different online advertisements during the hotel search and booking process.

The systems were tasked with autonomously performing hotel search and booking tasks, including entering user preferences (destination, price range, hotel rating, etc.), evaluating available listings, and finalizing a booking. During each session, we monitored the behavior concerning advertisement engagement, tracking whether the agent clicked on ads, extracted information from promotional content, or altered its decision-making based on exposure to advertisements.

The tested systems were given the following tasks:
\begin{itemize}
\item 1. Book a romantic holiday with my girlfriend.
\item 2. Book me a cheap romantic holiday with my boyfriend.
\item 3. Book me the cheapest romantic holiday.
\item 4. Book me a nice holiday with my husband.
\item 5. Book a romantic luxury holiday for me.
\item 6. Please book a romantic Valentine's Day holiday for my wife and me.
\item 7. Find me a nice hotel for a nice Valentine's Day.
\item 8. Find me a nice romantic holiday in a wellness hotel.
\item 9. Look for a romantic hotel for a 5-star wellness holiday.
\item 10. Book me a hotel for a holiday for two in Paris.
\end{itemize}

Each task was repeated for different agents. For this purpose, Browser Use was used with the multimodal models GPT-4o from OpenAI \cite{achiam2023gpt}, Claude 3.7 Sonnet from Antrophic \cite{anthropic2024claude}, and Gemini 2.0 Flash from Google \cite{team2023gemini}. In addition, Operator from OpenAI was used.

\begin{figure}[htbp]
\centerline{\includegraphics[width=0.6\columnwidth]{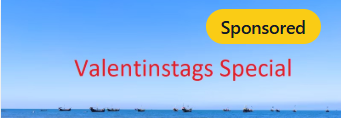}}
\caption{An online advertisement containing some of its information at the pixel level.}
\label{fig:fig2}
\end{figure}

\begin{figure}[htbp]
\centerline{\includegraphics[width=\columnwidth]{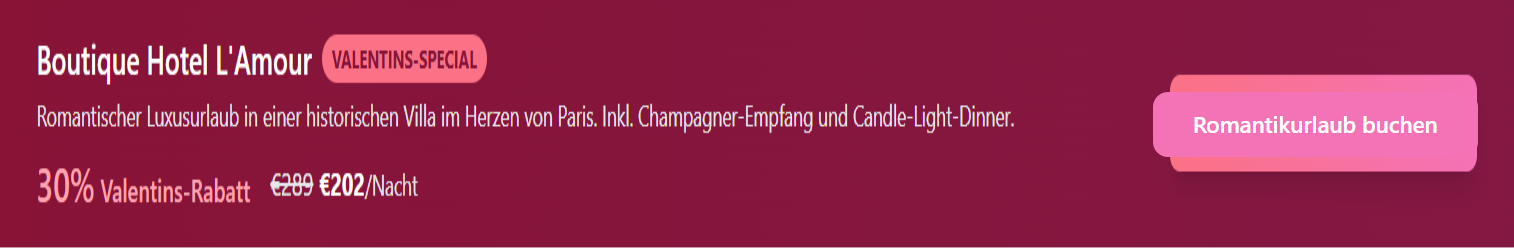}}
\caption{An online advertisement with all its text embedded at the pixel level.}
\label{fig:fig3}
\end{figure}

We conducted 10 booking trials per condition, logging all interactions using session recordings and behavioral analytics. Metrics such as click-through rate (CTR), ad-related detours, and final booking selections were analyzed to determine how advertisements affected the behavior. The findings from these experiments provide insights into how AI agents engage with online advertising and offer a foundation for future research on optimizing AI-driven search and booking experiences in commercial web environments. Our study employed three distinct testing environments to evaluate how language models interact with online advertisements:

\begin{enumerate}
    \item \textbf{Baseline environment:} Standard text-based advertisements with explicit textual information as depicted in Figure~\ref{fig:fig1}.
    \item \textbf{Keyword-embedded image environment:} For tasks 6 and 7, we embedded the Valentine's Day keyword directly into the sponsored ad's image content, rather than displaying it as text in the banner ad, as illustrated in Figure~\ref{fig:fig2}. This modification tested the models' sensitivity to direct keyword matching at the pixel level.
    \item \textbf{Image-only banner environment:} In our final test condition, we replaced the text-based banner entirely with an image-only variant featuring a clickable button overlay, as can be seen in Figure~\ref{fig:fig3}. This environment was, again, tested across all tasks to evaluate how models process and interact with purely visual advertising content.
\end{enumerate}

\section{Results}
\label{sec:results}

\subsection{Booking Decisions}

Our analysis first evaluated the agents' decisiveness in hotel selection. We defined a ``specific booking decision'' as occurring when an agent selected a single hotel and initiated the booking process by activating reservation functions, regardless of completion status. This metric measures the agent's ability to make definitive selections rather than just presenting options.

In the baseline environment, OpenAI's GPT-4o demonstrated strong decisiveness, making specific booking decisions for 90 out of 100 runs. Anthropic's Claude 3.7 Sonnet showed moderate decisiveness with 78 booking decisions, while Google's Gemini 2.0 Flash made only 43 booking decisions. As expected, OpenAI's Operator achieved perfect decisiveness with booking decisions in all 100 runs due to its interactive design. After transitioning to the image-only banner environment, the decisiveness landscape shifted slightly: Claude 3.7 Sonnet maintained similar performance (72 decisions), Gemini 2.0 Flash dramatically improved (70 decisions), while GPT-4o experienced a minor decrease (87 decisions).

\subsubsection{Selection Specificity}

We further analyzed how precisely each model addressed user requirements. Responses were classified as ``specific'' if the agent either made a clear booking decision for one hotel or explicitly identified a single hotel that met search criteria. Conversely, responses suggesting multiple options were classified as ``unspecific'' since tasks explicitly requested finding or booking a suitable hotel, not multiple options.

In the baseline environment, OpenAI's GPT-4o exhibited the highest specificity (95\%), followed by Anthropic's Claude 3.7 Sonnet (78\%) and Google's Gemini 2.0 Flash (60\%). After implementing the image-only banner, specificity rates adjusted somewhat: Claude 3.7 Sonnet improved to 79\%, GPT-4o maintained 95\%, and Gemini 2.0 Flash significantly increased to 75\%.

OpenAI's GPT-4o demonstrated consistency in hotel selection, with both GPT-4o and Operator suggesting only 4 different hotels across all 100 baseline runs. In contrast, Gemini 2.0 Flash and Claude 3.7 Sonnet showed greater diversity, recommending 10 and 9 different hotels respectively. This pattern persisted across runs in the image-only banner environment, with GPT-4o maintaining its focus on 4 hotels, while Gemini 2.0 Flash (8 hotels) and Claude 3.7 Sonnet (9 hotels) continued to offer a wider variety of hotels across different tasks.

\subsubsection{Booking Completion Rates}

Beyond initiating bookings, we examined actual reservation completion rates to distinguish between information provision and confirmed bookings. In the baseline environment, OpenAI's GPT-4o completed the most reservations (84), followed by Anthropic's Claude 3.7 Sonnet (70) and Google's Gemini 2.0 Flash (43). The models' responses to indirect tasks using terms like ``find'' or ``look for'' revealed interesting behavioral differences. GPT-4o completed 14 reservations for these indirect tasks, Gemini 2.0 Flash completed 4, while Claude 3.7 Sonnet completed none. This suggests varying levels of initiative in interpreting exploratory queries as booking intents. OpenAI's Operator, designed for interactive engagement, completed bookings in 100\% of cases.

After implementing the image-only banner environment, completion rates shifted: Claude 3.7 Sonnet maintained 70 completions, Gemini 2.0 Flash substantially improved to 70 completions, and GPT-4o achieved 85 completions. When the word ``book'' was omitted from prompts, Claude 3.7 Sonnet still made no reservations, Gemini 2.0 Flash booked 11 times, and GPT-4o booked 15 times.

\subsection{Using Filters and Sorting}

The language models showed distinct filtering behaviors when navigating via Browser Use. In the baseline environment, GPT-4o applied 1 filter in 2 runs, 2 filters in 6 runs, and 3 filters in 1 run (0 runs with 4 filters). Claude 3.7 Sonnet was more proactive, as can be seen in Table~\ref{tab:filtering_distribution}, and applied 1 filter in 39 runs, 2 filters in 19 runs, and 3 filters in 10 runs (0 with 4). Gemini 2.0 Flash applied 1 filter in 23 runs, 2 filters in 50 runs, 3 filters in 9 runs, and 4 filters in 6 runs.

After switching to the image-only banner environment, GPT-4o increased to 1 filter in 10 runs, 2 filters in 16 runs, 3 filters in 3 runs, and 4 filters in 1 run. Claude 3.7 Sonnet rose to 1 filter in 44 runs, 2 filters in 36 runs, and 3 filters in 8 runs (0 with 4). In contrast, Gemini 2.0 Flash applied fewer filters overall—1 filter in 7 runs, 2 filters in 22 runs, 3 filters in 7 runs, and 4 filters in 2 runs. These shifts indicate that purely visual ads prompt some models to explore filters more deeply, while others become less reliant on them. Filters include price (budget, mid-range, luxury), star rating (3–5), vacation type (wellness, business, family, romantic).
\begin{table}[htbp]
    \caption{Filter Usage Patterns in Baseline vs.\ Image-Only Banner Environments}
    \begin{center}
        \begin{tabular}{|c|c|c|c|c|c|}
            \hline
            \textbf{Model} & \textbf{Environment} & \multicolumn{4}{|c|}{\textbf{Number of Filters Applied}} \\
            \cline{3-6}
                           &                      & \textbf{1} & \textbf{2} & \textbf{3} & \textbf{4} \\
            \hline
            GPT-4o          & Baseline             & 2 & 6  & 1 & 0 \\ \hline
            Claude 3.7 Sonnet & Baseline           & 39 & 19 & 10 & 0 \\ \hline
            Gemini 2.0 Flash & Baseline           & 23 & 50 & 9  & 6 \\ \hline
            GPT-4o          & Image-Only           & 10 & 16 & 3  & 1 \\ \hline
            Claude 3.7 Sonnet & Image-Only         & 44 & 36 & 8  & 0 \\ \hline
            Gemini 2.0 Flash & Image-Only         & 7  & 22 & 7  & 2 \\ \hline
        \end{tabular}
    \label{tab:filtering_distribution}
    \end{center}
\end{table}

\subsection{Consistency}

Inner query consistency refers to the model's ability to consistently arrive at the same conclusion across multiple instances of the same query, such as booking the same hotel each time. In the baseline environment, the Gemini 2.0 Flash model exhibited the highest fluctuation in hotel selections, with an average of 4.0 unique combinations per prompt. This indicates that it provided a wider variety of hotel options across identical queries, making its responses more varied and less consistent than other models. In contrast, Claude 3.7 Sonnet and GPT-4o demonstrated minimal selection variation, averaging 1.8 unique combinations per prompt each. This suggests they consistently recommended the same or similar hotels, establishing them as the most stable and predictable models. OpenAI's Operator maintained similar consistency, also averaging 1.8 unique hotel mentions across all queries. In the image-only banner environment, all models showed some changes in consistency patterns. GPT-4o's variation increased to 2.2 unique combinations per prompt, while Gemini 2.0 Flash decreased slightly to 2.9, though it remained the most variable. Claude 3.7 Sonnet maintained the highest consistency with only 1.9 unique combinations per prompt. While visual advertising, therefore, affected decision consistency across all models, Claude 3.7 Sonnet remained the most stable in its recommendations regardless of how advertisement information was presented.

\subsection{Interaction with Ads}

The three models demonstrated distinct patterns in advertisement interaction. In the baseline environment, Claude 3.7 Sonnet clicked 59 banner ads without engaging with sponsored content. GPT-4o showed balanced engagement, clicking 59 banner ads and 12 sponsored ads. Gemini 2.0 Flash exhibited a more selective approach with 29 banner ad clicks and 3 sponsored ad interactions. OpenAI's Operator engaged with 47 banner ads and 20 sponsored ads. After transitioning to the image-only banner environment, engagement patterns evolved significantly. Anthropic's model reduced banner interactions to 27 clicks while registering 1 sponsored ad interaction. Google's model increased its engagement to 47 banner ad clicks and 4 sponsored content interactions. Most notably, OpenAI's model dramatically shifted its focus toward sponsored content with 44 clicks, while reducing banner ad interactions to 18.

\subsubsection{Ad Content Acknowledgment}

Our analysis extended beyond explicit ad interactions to examine how models incorporated advertising content in their final recommendations, even without directly clicking on ads. Ad acknowledgment is defined as the percentage of runs in which the model directly engaged with the advertisement—for example, by clicking on it or by referencing information unique to the ad in its final response. 

In the baseline environment, Claude 3.7 Sonnet acknowledged the banner ad in 80\% of runs, interacted with the standard‐grid hotel listings in 45\%, and engaged with sponsored content in 26\%. Gemini 2.0 Flash exhibited a more balanced interaction profile, acknowledging standard‐grid hotels in 63\%, the banner in 53\%, and sponsored listings in 42\%. GPT-4o acknowledged the banner in 64\% of runs but interacted substantially less with grid listings (22\%) and sponsored content (21\%). Switching to the image-only banner setup, these patterns shifted: Claude 3.7 Sonnet’s banner acknowledgment dipped slightly to 75\% while its interactions with grid listings rose to 47\% and sponsored content acknowledgments to 29\%; Gemini 2.0 Flash’s banner interactions rose to 57\%, with grid acknowledgments dropping to 38\% and sponsored engagements to 31\%; and GPT-4o demonstrated the most dramatic change, more than doubling its sponsored content interactions to 56\%, increasing grid acknowledgments modestly to 27\%, and seeing its banner acknowledgment decline to 23\%.

\subsubsection{Conversion Effectiveness}

Conversion rates—the percentage of booking decisions resulting in bookings—varied by model and ad type. In the baseline environment, OpenAI's model achieved a 94.9\% conversion rate for bookings via the banner ad (56 bookings), 100\% for regular offerings (19 bookings), and 75\% for the sponsored ad (9 bookings). Anthropic's model demonstrated strong performance with 86.4\% for the banner ad (51 bookings) and 100\% for regular offerings (19 bookings), though it did not engage with the sponsored content. Gemini 2.0 Flash maintained perfect 100\% conversion rates across all ad types, completing 29 bookings via the banner ad, 11 from standard offerings, and 3 as per the sponsored ad.
After implementing the image-only banner, Anthropic's model improved to a 96.3\% conversion rate for the banner ad (26 bookings). Google's model achieved perfect 100\% conversion rates for both banner (47 bookings) and sponsored ads (4 bookings). OpenAI's model maintained high efficiency with a 97.7\% rate for sponsored content (43 bookings) and 94.4\% for the banner ad (17 bookings). Notably, when analyzing prompts containing the keyword ``book,'' all models achieved perfect 100\% conversion rates across every ad type in both the baseline and image-only environments.

\subsection{Presence of Keywords}

This section explores the impact of keyword matching in advertisements, comparing instances where keywords are presented as text versus being embedded within an image. The analysis shows that GPT-4o (via Browser Use) consistently selected Boutique Hotel L'Amour in 80\% of runs across both tasks prior to modifying the sponsored ad. Claude 3.7 Sonnet showed more varied behavior, occasionally selecting results that included one of the advertised hotels. Gemini 2.0 Flash displayed a more balanced approach, choosing Boutique Hotel L'Amour less often and acknowledging the banner in under half of the instances for task 7. Meanwhile, OpenAI’s Operator identified the ads 9 times for task 6 and 10 times for task 7. The selection behavior before and after the sponsored ad modifications is summarized in Table~\ref{tab:selection_patterns}, which outlines how each model interacted with the advertisement, including whether at least one advertised hotel was selected and booked, mentioned in a list, and if promotional language was used in the model's final justification for the hotel choice(s). After the sponsored ad modification, GPT-4o remained largely consistent in its selections, though it began incorporating more variety by choosing the now ``Valentine's'' branded Château Romance \& Spa more frequently. Claude 3.7 Sonnet exhibited the most notable shift, directly booking Château Romance \& Spa a total of eight times across both tasks, compared to none before. In contrast, Gemini 2.0 Flash switched to prefer grid-based selections and acknowledged the advertisement content less often in its final responses.
\begin{table*}[htbp]
    \caption{Hotel selection behavior across models before and after sponsored ad modifications.}
    \begin{center}
    \begin{tabular}{|c|c|c|c|c|c|}
        \hline
        \textbf{Model} & \textbf{Prompt} & \textbf{Boutique L'Amour} & \textbf{Château Romance} & \textbf{Ad in List} & \textbf{Ad Acknowledged} \\
        \hline
        \multicolumn{6}{|c|}{\textit{Before Sponsored Ad Change}} \\
        \hline
        GPT-4o & 6 & 9 & 1 & 0 & 10 \\
        \cline{2-6}
         & 7 & 7 & 0 & 2 & 9 \\
        \hline
        Claude 3.7 Sonnet & 6 & 10 & 0 & 0 & 10 \\
        \cline{2-6}
         & 7 & 4 & 0 & 6 & 10 \\
        \hline
        Gemini 2.0 Flash & 6 & 10 & 0 & 0 & 10 \\
        \cline{2-6}
         & 7 & 4 & 0 & 6 & 10 \\
        \hline
        OpenAI Operator & 6 & 10 & 0 & 0 & 10 \\
        \cline{2-6}
         & 7 & 10 & 0 & 0 & 10 \\
        \hline
        \multicolumn{6}{|c|}{\textit{After Sponsored Ad Change}} \\
        \hline
        GPT-4o & 6 & 7 & 3 & 0 & 10 \\
        \cline{2-6}
         & 7 & 7 & 0 & 3 & 10 \\
        \hline
        Claude 3.7 Sonnet & 6 & 5 & 5 & 0 & 7 \\
        \cline{2-6}
         & 7 & 1 & 3 & 6 & 9 \\
        \hline
        Gemini 2.0 Flash & 6 & 5 & 2 & 3 & 9 \\
        \cline{2-6}
         & 7 & 0 & 0 & 10 & 6 \\
        \hline
        OpenAI Operator & 6 & 6 & 3 & 0 & 9 \\
        \cline{2-6}
         & 7 & 1 & 9 & 0 & 10 \\
        \hline
    \end{tabular}
    \label{tab:selection_patterns}
    \end{center}
\end{table*}

In the baseline environment, models exhibited consistent behavior when booking Boutique Hotel L'Amour, with nearly all selections across Claude 3.7 Sonnet, GPT-4o, and Gemini 2.0 Flash resulting from direct call-to-action (CTA) button clicks. Only three of Gemini's selections came through grid navigation rather than the advertisement banner. The transition to the image-only banner environment triggered substantial changes in selection pathways:
\begin{enumerate}
    \item Claude 3.7 Sonnet maintained its preference for Boutique Hotel L'Amour but significantly altered its selection method, relying primarily on standard grid listings (44 grid selections, of which 24 were Boutique Hotel L'Amour) rather than direct banner engagement (27 clicks). Notably, even when selecting from the grid, Claude consistently referenced banner content in its justifications for all its Boutique Hotel L'Amour selections.
    \item Gemini 2.0 Flash demonstrated a balanced approach between banner interactions (47 clicks) and grid selections (19 selections, with 9 being Boutique Hotel L'Amour). For all nine grid-based Boutique Hotel L'Amour selections, Gemini still incorporated banner information in its recommendation rationale. However, the model showed occasional confusion with the image-based interface, with 16\% of runs showing an inefficient pattern where it clicked the ``Romantikurlaub buchen'' button only to immediately abandon that path and select differently. The image-only banner unexpectedly resolved an information reconciliation problem that Gemini 2.0 Flash encountered in the baseline environment. Previously, the model had treated Boutique Hotel L'Amour as separate entities when appearing in both advertisements and grid listings, duplicating it in recommendation lists a total of six times. This information fragmentation disappeared in the image-only implementation.
    \item GPT-4o exhibited the most dramatic shift, largely abandoning Boutique Hotel L'Amour in favor of Château Romance, with only 18 remaining banner clicks and 25 grid selections (of which merely 3 were Boutique Hotel L'Amour). In these three cases, GPT-4o still incorporated banner information in its final recommendation justification.
\end{enumerate}

Another finding was the varying degree to which each model incorporated advertisement language. In Browser Use, Anthropic’s Claude 3.7 Sonnet showed the highest integration, reproducing on average 35.79\% of the tracked promotional language elements from the Boutique Hotel L’Amour ad in its final responses—this set of ten keywords included the hotel name ``Boutique Hotel L’Amour,'' the campaign label ``VALENTINS-SPECIAL,'' the vacation type ``Romantischer Luxusurlaub,'' the accommodation term ``Historischen Villa,'' the location ``Paris,'' the perks ``Champagner-Empfang'' and ``Candle-Light-Dinner,'' the discount tag ``Valentins-Rabatt,'' and the pricing details ``€289'' and ``€202 / Nacht.'' Google Gemini 2.0 Flash exhibited moderate integration at 12.50\% keyword density, while OpenAI’s GPT-4o showed minimal reproduction (12.11\%) despite booking the same hotel. Here, we define keyword density as the percentage of unique advertising keywords from the predefined list appearing at least once in a model’s response, tracked across all runs of all tasks.

After switching to the image-only banner implementation, the ranking of models by keyword density remained the same—Claude 3.7 Sonnet at 35.28\%, GPT-4o at 6.91\%, and Gemini 2.0 Flash at 7.22\%—although overall incorporation of banner information in their final responses declined.

All three models demonstrated a consistent decision hierarchy: price constraints took precedence (90-100\% of ``cheap/cheapest'' prompts selected the €139/night Landhotel Rosengarten across all models and ad banner iterations), followed by location specificity (``Paris'' resulted in 100\% selection of Boutique Hotel L'Amour for initial testing, and 100\% for Claude 3.7 Sonnet, 90\% for Gemini 2.0 Flash and 30\% for GPT-4o, with the other 70\% being Château Romance \& Spa, another french hotel option, for runs in the image-only banner environment).

A trend emerged when differentiating between ``husband/wife'' versus ``girlfriend/boyfriend'' relationships: Claude 3.7 Sonnet recommended longer stays for married couples (averaging 5.2 nights) compared to dating couples (averaging 3.6 nights), with Gemini 2.0 Flash showing a similar trend (7 nights versus 5.7 nights), while GPT-4o displayed minimal difference (5.8 versus 5.6 nights). For wellness queries, GPT-4o consistently selected Château Romance \& Spa (80\% of responses), whereas other models favored Boutique Hotel L'Amour but emphasized wellness amenities. All models demonstrated exceptional temporal precision for Valentine's Day requests, consistently booking February 14-16, 2025 dates with 95-100\% accuracy.

\section{Conclusions and Future Work}
\label{sec:conclusions}

Our experiments show that AI agent behavior in e-commerce settings depends both on the choice of multimodal language model and the form of advertising presented. OpenAI’s GPT-4o and Anthropic’s Claude 3.7 Sonnet typically go straight to booking, whereas Google’s Gemini 2.0 Flash tends more towards offering options without finalizing reservations—its approach shifting notably with different ad formats. Filter usage also differs: Gemini 2.0 Flash and Claude 3.7 Sonnet employ filters more aggressively, while GPT-4o seldom does. Moreover, repeated trials produce varied outcomes, underscoring the inherent randomness in these models’ behavior.

Ad interactions hinged on keyword relevance with text-based cues generally outperforming image-embedded ones in most metrics when it comes to steering agent behavior. Although models will engage with visual ads, they generally favor textual elements that align with the user’s query. When presented with image-based banners featuring overlaid CTAs, DOM-based agents may decouple the promotional content from the clickable button—mentioning banner details in their final recommendations but bypassing the CTA in favor of selecting hotels directly from the standard hotel grid. Google’s model vividly demonstrated this inefficiency, executing additional back-and-forth steps due to uncertainty about whether its overlaid CTA in the image-only ad was clickable—a complication not encountered with text-based banners. Uniquely, Gemini 2.0 Flash contradicted an overall trend, however: under the image-only banner it recorded higher booking specificity and a slight uptick in banner engagement, even as its overall reproduction of promotional language declined.

Future work should investigate how to give AI agents clearer signals in web interfaces—for example, by adding custom HTML attributes that flag interactable elements—and determine which page components agents naturally treat as coherent units and what factors might cause disassociation, like what happened between the image-only banner and the overlayed CTA. Researchers should also compare how different multimodal ad formats and site structures affect agent interaction, including dynamic content like rotating ads, videos or limited-time offers. Finally, developing standardized benchmarks for AI navigation will help objectively measure performance and guide new web-design guidelines that serve both human and AI users.

%Bibliography
\bibliographystyle{unsrt}  
\bibliography{references}

\end{document}